\documentclass[sigconf, 
nonacm %preprint
]{acmart}

\usepackage{subcaption}
\usepackage[ruled,noend]{algorithm2e}
\usepackage[T1]{fontenc}
\usepackage[utf8]{inputenc}

\usepackage{stfloats}
\usepackage{booktabs}

\usepackage{array}
\newcolumntype{L}[1]{>{\raggedright\let\newline\\\arraybackslash\hspace{0pt}}p{#1}}
\newcolumntype{C}[1]{>{\centering\let\newline\\\arraybackslash\hspace{0pt}}p{#1}}
\newcolumntype{R}[1]{>{\raggedleft\let\newline\\\arraybackslash\hspace{0pt}}p{#1}}

\usepackage{tikz}
\usetikzlibrary{
    arrows.meta,
    chains,
    positioning,
    shapes.geometric
}

\pgfdeclarelayer{background}
\pgfdeclarelayer{foreground}
\pgfsetlayers{background,main,foreground}

\tikzset{
    FlowChart/.style={
        startstop/.style = {
            rectangle, 
            draw,
            minimum width=3cm, 
            minimum height=0.8cm,
            on chain, 
            join=by arrow
        },
        io/.style = {
            rectangle, 
            draw,
            minimum width=1.8cm, 
            minimum height=0.5cm,
        },
        set/.style = {
            rectangle, 
            draw, 
            double,
            double distance=1mm,
            minimum width=3cm, 
            text width=4cm,
            minimum height=0.9cm, 
            on chain, 
            join=by arrow,
            align=center
        },
        process/.style = {
            rectangle, 
            rounded corners, 
            draw,
            minimum width=3cm,
            minimum height=0.8cm, 
            align=center,
            on chain, 
            join=by arrow
        },
        process2/.style = {
            rectangle, 
            rounded corners, 
            draw,
            text width=3cm, 
            minimum height=0.8cm, 
            align=center,
        },
        decision/.style = {
            diamond,
            aspect=1.7,
            draw,
            minimum width=2.5cm, 
            minimum height=1cm, 
            align=center,
            on chain, 
            join=by arrow
        },
        decision2/.style = {
            diamond,
            aspect=1.7,
            draw,
            minimum width=2.5cm, 
            minimum height=1cm, 
            align=center,
        },
        arrow/.style = {
            thick,
            -Triangle
        },
    }   
}

\usepackage{listings}
\usepackage{color}

\definecolor{dkgreen}{rgb}{0,0.6,0}
\definecolor{gray}{rgb}{0.5,0.5,0.5}
\definecolor{lightgray}{RGB}{239,240,241}
\definecolor{mauve}{rgb}{0.58,0,0.82}
\definecolor{lightgreen}{rgb}{0.8,1,0.8}
\definecolor{lightyellow}{rgb}{1,1,0.8}
\definecolor{lightred}{rgb}{1,0.8,0.8}

\usepackage[norndcorners,customcolors]{hf-tikz}
\hfsetbordercolor{lightgray}
\hfsetfillcolor{lightgray}

\AtBeginDocument{%
  \providecommand\BibTeX{{%
    \normalfont B\kern-0.5em{\scshape i\kern-0.25em b}\kern-0.8em\TeX}}}

\begin{document}

% from https://gecco-2020.sigevo.org/index.html/Papers+Submission+Instructions
\acmDOI{10.1145/nnnnnnn.nnnnnnn} % To be updated after completing copyright process
\acmISBN{978-x-xxxx-xxxx-x/YY/MM} % To be updated after completing copyright process
\acmConference[GECCO '20]{The Genetic and Evolutionary Computation Conference 2020}{July 8--12, 2020}{Cancun, Mexico}
\acmYear{2020}
\copyrightyear{2020}

\title{Optimising the Fit of Stack Overflow Code Snippets into Existing Code}

\author{Brittany Reid}
\email{brittany.reid@adelaide.edu.au}
\affiliation{%
  \institution{University of Adelaide}
  \country{Australia}
}

\author{Christoph Treude}
\email{christoph.treude@adelaide.edu.au}
\affiliation{%
  \institution{University of Adelaide}
  \country{Australia}
}

\author{Markus Wagner}
\email{markus.wagner@adelaide.edu.au}
\affiliation{%
  \institution{University of Adelaide}
  \country{Australia}
}

\sloppy

\begin{abstract}
Software developers often reuse code from online sources such as Stack Overflow within their projects. However, the process of searching for code snippets and integrating them within existing source code can be tedious. In order to improve efficiency and reduce time spent on code reuse, we present an automated code reuse tool for the Eclipse IDE (Integrated Developer Environment), NLP2TestableCode. NLP2TestableCode can not only search for Java code snippets using natural language tasks, but also evaluate code snippets based on a user's existing code, modify snippets to improve fit and correct errors, before presenting the user with the best snippet, all without leaving the editor. NLP2TestableCode also includes functionality to automatically generate customisable test cases and suggest argument and return types, in order to further evaluate code snippets. In evaluation, NLP2TestableCode was capable of finding compilable code snippets for 82.9\% of tasks, and testable code snippets for 42.9\%.

\end{abstract}




\maketitle

\section{Introduction}
Among software developers, reusing code snippets from the Internet is a common occurrence, with 79\% of developers reporting that they copied code from the popular programming question and answer site Stack Overflow (SO) \cite{SO} for use in their own projects within the last month \cite{Baltes2019}. While the benefits of this kind of code reuse are hard to quantify, case studies have previously observed a return of investment of up to 400\% \cite{wclim}. With 19 million questions and 29 million answers as of March 2020 \cite{SEstats}, Stack Overflow is one of the most popular resources for code snippets; however, due to their crowd-sourced nature, the quality of code snippets varies, with only 8.41\% of answers containing compilable code \cite{terragni_liu_cheung_2016}. This makes the process of integrating code snippets time-consuming; time that could be spent writing code is instead spent correcting compiler errors.

Existing code reuse tools like NLP2Code \cite{campbell_treude_2017} and Blueprint \cite{Brandt:2010:EPI:1753326.1753402} automate the process of searching for code snippets within the editor; however, snippets are inserted as-is and developers must still make changes in order to integrate them into existing source code. On the other hand, tools like CSNIPPEX \cite{terragni_liu_cheung_2016} and Jigsaw \cite{Cottrell:2008:JTS:1370175.1370194} automate code corrections and integration, but rely on a developer to supply a code snippet. As it stands, no existing tool attempts to automate the entire code reuse process, nor assist developers with testing code from online sources.

\begin{figure}[h]
\centering
%Begin flowcart
\begin{tikzpicture}[FlowChart, node distance = 3.7mm and 3mm, start chain = A going below, scale=0.75, every node/.style={scale=0.75}]

%outer
\node[startstop](A1) {Natural Language Task};
\node[process](A2) {Search Stack Overflow Database};
\node[set](A3) {Set of retrieved Code Snippets};

%inner
\node[process, below = 0.7](B1){Compile};
\node[decision](B2){Has Errors?};
\node[process](B3){Add to Set of Processed Snippets};

\path (B1.north) +(-1.3,+0.2) node (label1) {For Each Snippet:};

%inner box
\begin{pgfonlayer}{background}
    \path (B1.west |- B1.north)+(-1.02,0.5) node (a) {};
    \path (B3.east |- B3.south)+(+0.3,-0.5) node (b) {};
    \path[fill=yellow!20,rounded corners, draw=black!50, dashed](a) rectangle (b); 
\end{pgfonlayer}

%errors
\path(B2.east)+(3.6, 3.585) node (C1)[process2]{Integrate};
\path(C1.south)+(0,-1.3) node (C2) [decision2]{Has Errors?};
\path(C2.south)+(0,-1) node (C3) [process2]{Targeted Fixes};
\path(C3.south)+(0,-1.38) node (C4) [decision2]{Has Errors?};
\path(C4.south)+(0,-1) node (C5) [process2]{Line Deletion};

\path (C1.north) +(-0.7,+0.2) node (label2) {Code Correction:};

%error box
\begin{pgfonlayer}{background}
    \path(C1.west |- C1.north)+(-0.3,0.5) node (c){};
    \path(C5.east |- C5.south)+(+0.3,-0.5) node (d){};
    \path[fill=yellow!20,rounded corners, draw=black!50, dashed](c) rectangle (d); 
\end{pgfonlayer}

%arrows
\draw[arrow](B2.east) node[above right]{Yes} -- + (1.15,0) |- (C1);
\draw[arrow](C1.south) -- (C2.north);
\draw[arrow](C2.south) node[below right]{Yes} -- (C3.north);
\draw[arrow](C3.south) -- (C4.north);
\draw[arrow](C4.south) node[below right]{Yes} -- (C5.north);
\draw[arrow](C5.west) -| ++(-5.05mm,0mm) |- (B3.east);
\draw[arrow](C2.west) node[above left]{No} -| ++(-8.6mm,0mm) |- (B3.east);
\draw[arrow](C4.west) node[above left]{No} -- (B3.east);
%label some arrows
\node[below right] at (B2.south){No};

%outer
\node[process, below=0.7](A4) {Sort Processed Snippets by Quality};
\node[startstop](A5){Best Snippet};

\end{tikzpicture}
%End Flowchart
\caption{NLP2TestableCode’s process, from task to snippet.}
\label{fig:flowchart}
\end{figure}
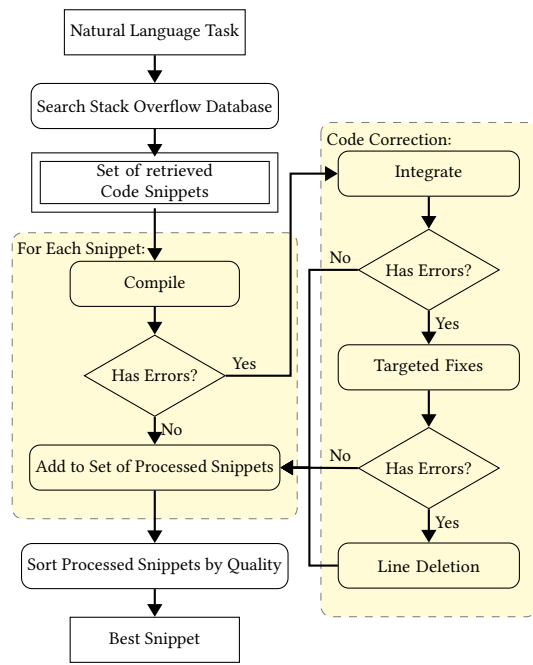

To address these issues, we employ a combination of methods from data-driven search-based software engineering (DSE)~\cite{nair2018dse}, resulting in NLP2TestableCode, a plug-in for the Eclipse IDE that (1) uses natural language tasks to search for relevant Java code snippets from a database of SO threads, (2) integrates code snippets by making changes based on existing source code, (3) corrects compiler errors and (4) provides automated testing tools to find working snippets. By automating all of these individual parts of the code reuse process, NLP2TestableCode aims to improve productivity and free up developers for other work. Because context switching has been shown to have a negative effect on productivity \cite{Proksch2015build}, we implement NLP2TestableCode as an in-editor tool that reduces the need to switch between the IDE and web browser.

%want this at the top of current page so we need to put it here as latex is stubborn 
\begin{figure*}[t]
\begin{subfigure}[t]{0.48\linewidth}
\caption{Snippet}
\begin{center}
\includegraphics[width=0.5\linewidth]{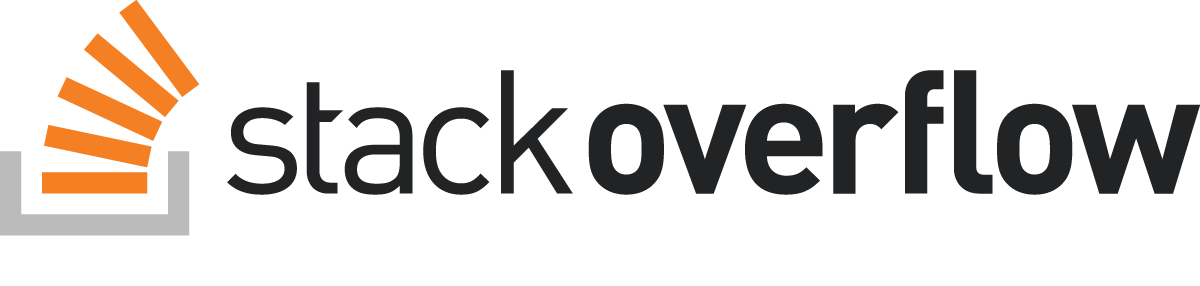}  \\
\end{center}
Alternatively, you can use an Ints method from the Guava library:
\begin{lstlisting}[aboveskip=2pt]
import com.google.common.primitives.Ints;
import java.util.Optional;
\end{lstlisting}
This makes for a concise way to convert a string into an int:
\begin{lstlisting}[aboveskip=2pt]
int foo = 0;
foo = Optional.ofNullable(myString)
     .map(Ints::tryParse)
     .orElse(0);
\end{lstlisting}
\label{subfig:examplesnip}
\end{subfigure}
\hfill
\begin{subfigure}[t]{0.48\linewidth}
\caption{Copy and Paste}
\begin{lstlisting}[backgroundcolor=\color{white}, aboveskip=2pt]
public class Main{
    public static void 
    main(String[] args){
\end{lstlisting}
\begin{lstlisting}[backgroundcolor=\color{lightgreen}]
        import com.google.common.primitives.Ints;
        import java.util.Optional;
            
        int foo = 0;
        foo = Optional.ofNullable(myString)
    		 .map(Ints::tryParse)
             .orElse(0);
\end{lstlisting}
\begin{lstlisting}[backgroundcolor=\color{white}]
    }
}
\end{lstlisting}
\label{subfig:examplecopy}
\end{subfigure}\vspace{-3mm}
\caption{Example code snippet adapted from a SO post \cite{example} and the result after being inserted into an existing file.}
\label{fig:example}
\end{figure*}

The process, from task to inserted snippet, can be seen in Figure~\ref{fig:flowchart}. Users begin by entering a natural language task where they would like to insert a snippet. Using this task, the plug-in will find relevant SO threads to extract code snippets from. For each code snippet, a version of the user's code with this snippet inserted is constructed then compiled. Snippets that successfully compile are added to the final set of processed snippets, while non-compilable snippets undergo the code correction process. Here, NLP2TestableCode attempts to integrate, correct specific compiler errors and delete lines from snippets to reduce the number of errors. When snippets compile, or are finished being processed, they are added to the set of processed snippets, which is sorted by number of errors. The first snippet, the one with the least compiler errors, will be inserted into the user's code. From here, users can optionally test the retrieved set of snippets, being provided with input and output type suggestions and a default JUnit test case to customise. The ordering of snippets is then updated based on number of passed tests, changing the inserted snippet if necessary.

\begin{table}[h]
  \caption{Comparison of NLP2Code and NLP2TestableCode.}
\begin{tabular}{@{}l R{1.39cm} R{1.67cm} R{1.55cm}@{}}
  \toprule
  \textbf{Plug-in} & \textbf{Snippets} \textbf{Retrieved} & \textbf{Tasks with Compilable Snippets} & \textbf{Tasks with Testable Snippets} \\
\midrule
NLP2Code & 355 & 21.3\% & 0\% \\
  NLP2TestableCode & 6,954 & 82.9\% & 42.5\% \\
  \bottomrule
\end{tabular}
  \label{tab:results}
\end{table}

We measured NLP2TestableCode against 47 tasks and compared these results to NLP2Code. A summary of this comparison is presented in Table \ref{tab:results}, NLP2TestableCode is capable of presenting users with compilable code snippets for 82.9\% of sample tasks, while increasing the number of compilable snippets out of the total retrieved from 4.7\%  to 29.3\% snippets using code correction approaches.

The public GitHub repository for NLP2TestableCode is available at: \url{https://github.com/Brittany-Reid/nlp2testablecode}

\section{Motivating Example}
Consider a typical code reuse situation where a developer would like to find an example Java code snippet illustrating how to convert a string into an integer. First, the developer would need to search for snippets; in this case the developer enters the query "How to convert string to int in Java" into their search engine. The first Stack Overflow thread returned for this query has 44 answers, each containing a code snippet. A developer cannot insert, integrate and test every snippet within a reasonable time, instead they must rely on additional information such as votes and comments, or their own programming knowledge to select suitable snippets.

\begin{figure}[h]
\begin{lstlisting}[backgroundcolor=\color{lightyellow}, aboveskip=0.2cm]
    import com.google.common.primitives.Ints;
    import java.util.Optional;
\end{lstlisting}
\begin{lstlisting}
    public class Main {
	    public static void main(String[] args){
\end{lstlisting}
\begin{lstlisting}[backgroundcolor=\color{lightgreen}]
+		    String myString;
+		    myString = "empty";
\end{lstlisting}
\begin{lstlisting}
		    int foo = 0;
	    	foo = Optional.ofNullable(myString)
	    	    .map(Ints::tryParse)
	    	    .orElse(0);
	}
}
\end{lstlisting}
\caption{The snippet in Figure \ref{fig:example} modified to compile.}
\label{fig:example2}
\end{figure}
Figure~\ref{subfig:examplesnip} shows an example SO answer and the embedded snippet within. The first step to integrating this snippet is to copy and paste it into an existing file. The result of this copy and paste can be seen in Figure \ref{subfig:examplecopy}. In this state the file will not compile; the import statements are not in the correct place and the variable myString is missing a declaration. To correctly integrate this snippet, the developer must make a series of changes to correct these problems, including moving the import statement to the start of the file and inserting a declaration and definition for myString. The resulting compilable snippet can be seen in Figure \ref{fig:example2}.

In contrast, using NLP2TestableCode only requires a developer to enter their task within Eclipse, after which they will be presented with set of snippets modified for them and sorted by best fit. NLP2TestableCode is able to automatically move import statements, fix common syntax errors like missing semi-colons and add missing variable declarations.

\section{Related Work}
Similar tools that help developers find online code snippet within their IDE include NLP2Code \cite{campbell_treude_2017}, Blueprint \cite{Brandt:2010:EPI:1753326.1753402}, Seahawk \cite{Ponzanelli:2013:SSO:2486788.2486988}, Prompter \cite{ponzanelli2014prompter} and Bing Developer Assistant (BDA) \cite{bda}. There has also been much research into using neural networks to map natural language to code snippets \cite{YinP}\cite{StaQC}. While some in-editor tools attempt to evaluate the quality or fit of code snippets, such as NLP2Code's use of SO vote counts to rank snippets and Prompter's ranking system that takes into account existing code \cite{ponzanelli2014mining}, none focus on automating the entire code reuse process, including the integration of code snippets.

Research into automated code corrections on Stack Overflow snippets has found that the number of compilable Java snippets could be improved from 1\% to 3.02\% through simple code fixes, such as adding missing semi-colons \cite{Yang:2016:QUC:2901739.2901767}. Tools like Jigsaw \cite{Cottrell:2008:JTS:1370175.1370194} and CSNIPPEX \cite{terragni_liu_cheung_2016} explore approaches to automated code correction and integration. Jigsaw is a plug-in for Eclipse that can take a code snippet and modify it in order to integrate it within an existing project, while CSNIPPEX takes a Stack Overflow URL and attempts to generate a compilable file from the snippet by correcting compiler errors using Eclipse's Quick Fix functionality \cite{quickfix}. However, neither of these tools aid the user in finding snippets. 

$\mu$SCALPEL \cite{Barr:2015:AST:2771783.2771796} is an automatic code transplant tool that uses genetic programming and testing to transplant code from one program to another. Using test cases to define and maintain functionality, small changes are made to the transplanted code, and code that does not aid in passing tests can be discarded, reducing the code to its minimal functioning form. $\mu$SCALPEL is limited in that it is designed to transplant code from one working program to another, not example code like that found on Stack Overflow.

\section{Approach}
NLP2TestableCode undergoes a multistage process to take a natural language task and insert into the user's code the snippet that best fits. These steps include retrieving relevant code snippets from a database of Stack Overflow threads, evaluating code snippets within the context of a user's existing code, modifying snippets that do not compile through the use of integration, targeted fixes and line deletion and then finally sorting the processed set of snippets and inserting the best snippet. NLP2TestableCode is implemented as a plug-in for Eclipse that functions in a single window, and users can cycle through processed snippets as they become available. This cycling replaces the currently inserted snippet with the next best, allowing a user access to more than a single chosen snippet.

An optional testing stage is included to further evaluate snippets. When evaluation and correction is complete, compilable snippets are processed to determine possible argument and return types for testing. This type information is used to suggest test input and output, then used to construct both a customisable skeleton JUnit test case and testable functions from code snippets. After testing, snippet ordering is updated based on passed tests and the inserted snippet is updated with a new best if necessary.

\subsection{Using  Natural  Language  Tasks  to  Find  Relevant Snippets}

\begin{figure}[h]
\includegraphics[width=0.9\linewidth]{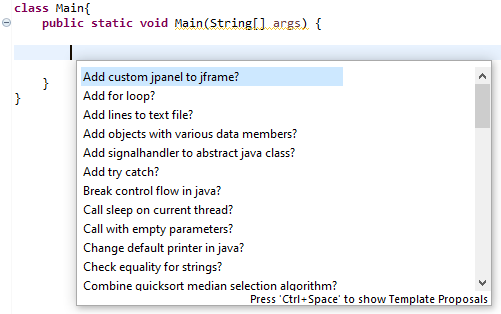}
\caption{Task suggestions through content assist.}
\label{fig:contentassist}
\end{figure}

Searching for code snippets effectively is a crucial aspect of automating the code reuse process. Fortunately, code snippets on Stack Overflow are surrounded by natural language questions, explanations and comments. The challenge of mapping tasks to code snippets is ensuring that as many relevant snippets are retrieved as possible.

Online searches like the one used in NLP2Code \cite{campbell_treude_2017}, with a hard-coded limit of 12 snippets per task, are restricted by the time required to download each Stack Overflow thread. To address this, NLP2TestableCode uses an offline SO database, pre-filtered to only include threads tagged with Java, and is capable of retrieving hundreds of snippets in less than a second. The size of this filtered dataset is 6.84GB, containing 1.5 million questions and 2.5 million answers. 

Stanford's CoreNLP~\cite{core} is used to lemmatise queries and question titles, reducing their words to common forms. Stop words like \textit{the}, \textit{an} and \textit{is} are removed, using a list from NLTK~\cite{nltk} for Python, along with the word \textit{java}, as threads are already filtered by language. Each word in a processed title is added to a database, associated with a list of threads with that word in the title; similarly, processed query words are used to retrieve those same sets of threads.

Users simply enter a task where they would like a snippet to be inserted, either by selecting a task from the suggestions presented through Eclipse's content assist feature, based on NLP2Code's \cite{campbell_treude_2017} task suggestion functionality, or by ending a custom task with a question mark. With the selected task, the plug-in will begin the process of searching, evaluating and fixing snippets. Combined, these features enable users to integrate code from SO into an existing file without having to leave the editor.

\subsection{Evaluating Code Quality}
NLP2TestableCode presents the user with the `best' snippet from the total set of retrieved snippets for a given task. Because the plug-in aims to reduce the amount of manual integration work, snippets should be evaluated based on how well they integrate within the user's existing code, with the snippet requiring the least work to integrate being presented to the user first. To do this, snippets and the user's existing code are combined at the point of task entry, then this code is compiled to count errors. An example of this combined code can be seen in Figure \ref{fig:maininmain}.

\begin{figure}[h]
\begin{lstlisting}[tabsize=4, aboveskip = 0.2cm]
class Main{
	public static void main(String[] args){
\end{lstlisting}
\begin{lstlisting}[backgroundcolor=\color{lightgreen}, tabsize=4]
+		public static void main(String[] args){
+			int result = Integer.parseInt(args[0]);
+		}
\end{lstlisting}
\begin{lstlisting}[tabsize=4]
	}
}
\end{lstlisting}
 \caption{A snippet, highlighted, inserted into existing code.}
 \label{fig:maininmain}
\end{figure}

By compiling snippets within the context of a user's existing source code, NLP2TestableCode can apply a context sensitive analysis of each snippet. The logic is that snippets which integrate well within existing code should produce less errors than those that do not. By ranking snippets by compiler errors, snippets that best fit existing code can be shown to the user before others. This also means that the set and ordering of compiling snippets can change based on insertion location and surrounding code. Likewise, using compiler errors as a measure of quality enables snippets containing syntax errors or missing elements to be ranked lower than those without errors. The snippet highlighted in Figure \ref{fig:maininmain} is an example of an otherwise correct snippet that contains elements that would cause it to fail to compile when inserted into an already existing main function. However, a snippet that contains only the inner statement would compile with no errors.

Because NLP2TestableCode preforms many compiles, it employs the use of an in-memory compiler to reduce compile time. By using in-memory compilation, no files are written to the disk during evaluation. The plug-in makes use of the Eclipse Compiler, the same compiler used within the IDE to underline compiler errors and warnings; because of this the Eclipse Compiler is better suited to compiling incomplete or incorrect code.

\subsection{Code Correction and Integration}
Most code snippets on Stack Overflow do not compile when inserted as-is. Using a range of automatic code correction techniques, NLP2TestableCode is able to improve the number of compilable snippets and reduce the amount of integration work required from users. All snippets that contain compiler errors are sent through the code correction process. Changes to snippets are only accepted if they decrease the number of compiler errors.

\subsubsection{Automatic Integration}
Snippets on Stack Overflow are written for example purposes and thus snippets range from single statements to full classes with multiple functions. Depending on the insertion location, these elements may be unnecessary and during manual integration be removed or shifted around. NLP2TestableCode is capable of handling a subset of these instances automatically, `snippetising' larger pieces of code.

Firstly, all snippets have any import statements extracted during initial processing. These import statements are stored separately from the rest of the snippet, to be inserted into their correct location when required. Without separating import statements, many snippets would have them inserted in incorrect locations.

Frequently SO answers wrap example code within potentially unnecessary classes and functions. Many of these class and method declarations can be removed without altering a snippet's functionality. Snippets are parsed to determine if they contain a class or function. Where snippets contain more than a single class and/or function, they are skipped; these are a more complex case the plug-in currently cannot handle. Classes that contain fields are also ignored, assuming that a snippet that includes fields is demonstrating their usage in some way, and that the class itself is part of the snippet's functionality.

NLP2TestableCode handles the simple case of inserting snippets that contain a main function into an existing main function. This is considered a simple case because the arguments of both functions will be the same. The function declaration can simply be removed along with the closing bracket. After the integration process has been preformed, snippets are compiled and the changes are only kept if they reduce compiler errors. This ensures that the integration process improves the correctness of a snippet.

\subsubsection{Targeted Fixes}

Many snippets on SO contain syntax errors, missing imports and undeclared variables that a developer would typically need to manually correct in order to integrate a snippet. NLP2TestableCode addresses these common compiler errors with targeted fixes, and can insert missing semi-colons and other tokens, find missing import statements, add variable declarations for undefined variables and remove error causing tokens. Compiling a snippet generates a list of diagnostic objects for each error, that contain error codes, location information and error messages. Using both error codes and information within error messages, targeted fixes can be applied to a snippet. The plug-in attempts to fix each error once, and if the fix reduces the number of errors, the changes to the snippet are accepted. Because one fix can sometimes resolve multiple errors and in order to avoid skipping any errors, previously processed errors must be stored and used to recalculate the next error to process.

The plug-in is capable of looking through packages on the Eclipse project classpath to solve missing import statements. It is not uncommon for type names to be used by multiple packages, which makes determining the correct one to use a challenge. In this case, the plug-in prefers classes that belong to packages in the default Java library. This allows common packages like \texttt{java.util.List} to be preferred over less common alternatives like \texttt{com.sun.tools.javac.util.List}. 

When undefined variables are found, the plug-in utilises JavaParser \cite{javaParser} to analyse usage and determine a type. For example, in the line of code \texttt{var = "some text";} the variable \texttt{var} can be assumed to be of type String by the value it is being assigned. This type information is then used to define and assign a default value to the variable. If no type can be determined through usage, the plug-in brute forces common types such as \texttt{Integer}, \texttt{Character}, \texttt{String}, \texttt{Boolean}, \texttt{Double}, \texttt{Long} and \texttt{Float}.

\subsubsection{Line Deletion}

NLP2TestableCode's final stage of code correction is line deletion. The aim of line deletion is to reduce a snippet into its optimal form through small changes. Line deletion uses a local search algorithm, detailed in Algorithm \ref{alg:del}. The current best, $S_{best}$, is initialised with the unmodified snippet. For each loop over the snippet, lines are deleted in order starting at the bottom of a snippet and each deletion is accepted if it does not increase the number of errors. A snippet is continuously looped over until no more changes can be made.

\begin{algorithm}
  $S_{best} \gets \text{Initial snippet}$\;
  $\text{done} \gets false$\;
  \While{$done == false$}{
     $\text{done} \gets true$\;
     $\text{line} \gets  S_{best}\text{.length}$\;
         \For{\text{int j =0  to } $S_{best}\text{.length} - 1$}{
          \If{$S_{best}(line)\text{.Deleted} $}{
          $line \gets line - 1$\;
                  \text{{\textbf{continue}}}
         }
         	$S_{current} \gets S_{best}$\; 
            $S_{current} \text{.}Delete(line)$\;
            $\text{errors} \gets  Compile(S_{current})$\;
             \If{$\text{errors } <= S_{best}$\text{.Errors}}{
              	$S_{best} \gets S_{current} $\;
                $\text{done} \gets false$\;
          }
            
            $line \gets line - 1$\;
         }
  }
    \KwRet{$S_{best}$}\;
  \caption{Deletion Algorithm (Local Search)}
  \label{alg:del}
\end{algorithm}

\subsection{Automated Testing}

NLP2TestableCode automates the testing process by integrating JUnit. After retrieving a set of snippets, a user can choose to test these snippets without leaving the current file. The plug-in provides recommendations for argument and return types, and using a given set of argument and return types can generate a default JUnit test case. This test case is inserted into the open file where a user can customise it, and with the press of a button use this test case to test the set of compiling snippets. After testing is complete, the inserted snippet updates with the new best. This testing process also requires transforming snippets into testable functions with input and output.

\subsubsection{Suggesting Argument and Return Types}
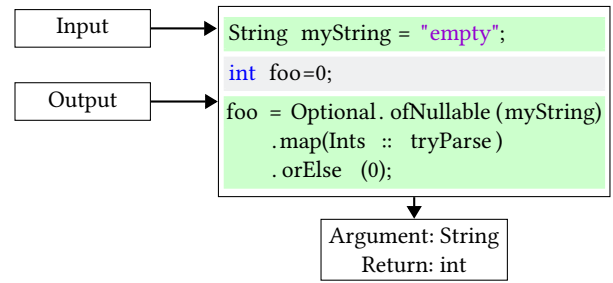
\begin{figure}[h]
\centering
%Begin flowcart
\begin{tikzpicture}[FlowChart, node distance = 5mm and 7mm, start chain = A going below, scale=1, every node/.style={scale=1}]

\node[io](A1){Output};
\path(A1.south)+(0, 1.23) node (A2)[io]{Input};
\path(A1.east)+(3.5, 0) node (A3)[io, align=left]{
\begin{lstlisting}[linewidth=5cm, backgroundcolor=\color{lightgreen}]
String myString = "empty";
\end{lstlisting} \\
\begin{lstlisting}[linewidth=5cm]
int foo=0;
\end{lstlisting} \\
\begin{lstlisting}[linewidth=5cm, backgroundcolor=\color{lightgreen}]
foo = Optional.ofNullable(myString)
    .map(Ints :: tryParse)
    .orElse (0);
\end{lstlisting}
};
\path(A3.south)+(0, -0.7) node (A4)[io, align=center]{Argument: String \\ Return: int};

\path (A3.west |- A3.north)+(0.1,-0.3) node (a) {};

\draw[arrow](A1.east) -- (A3.west);
\draw[arrow](A2.east) -- (a);
\draw[arrow](A3.south) -- (A4.north);

\end{tikzpicture}
\caption{How a snippet is processed for input and output.}
\label{fig:types}
\end{figure}

An important part of automating the testing process is being able to identify from a piece of code, which variables could be input and which output. By analysing compilable code snippets, the plug-in can attempt to guess appropriate input and output types. The plug-in does two things: it assumes that the last line of code in a snippet must be relevant to the functionality in some way, and analyses this line for a possible return argument, while it also looks at the variable declarations within a snippet to guess arguments. From these variables, the type information is extracted and used to provide suggestions for testing. This can be seen in the snippet in Figure \ref{fig:types}, the last variable being assigned, \texttt{foo}, will be chosen as output, while the variable myString will be selected as input. From these variables, the types \texttt{String} and \texttt{int} will be extracted, to be used as a type suggestion for a argument and return value.

If a user chooses to test a set of retrieved snippets, the return and argument type suggestions will be generated and displayed to the user. The user can also choose to insert their own type information. This type information, as well as the number of arguments, is used to construct both a testable function from a snippet, but also a skeleton JUnit test case. The test case's default input is generated using JavaParser's default type value information. An example JUnit test case for the types \texttt{String} and \texttt{int} can be seen in Figure \ref{fig:testcase}.

\begin{figure}[h]
\begin{lstlisting}[aboveskip = 0.2cm]
@Test
public void JUnitTest(){
    assertEquals(snippet("empty"), 0);
}
\end{lstlisting}
\caption{An automatically generated JUnit test case.}
\label{fig:testcase}
\end{figure}

\subsubsection{Building a Testable Function}
Informed by the provided argument and return type information, the plug-in attempts to generate from each snippet a function with input and output that can be tested. This process is similar to the one used to suggest argument and return types from variables, but with added information. Instead of looking at all variables, this search only looks for variables of the given type and number. The last variable of a specified type is accepted as the return, then variables that match argument types are searched for starting at the beginning of the snippet. Where not enough variables of the given types can be found, a testable function is unable to be generated. 

\begin{figure}[h]
\begin{lstlisting}[aboveskip = 0.2cm]
public static int snippet(String myString){
	int foo = 0;
    foo = Optional.ofNullable(myString)
 	    .map(Ints :: tryParse)
        .orElse (0);
        return foo;
}
\end{lstlisting}
\caption{Snippet from Figure \ref{fig:types} converted into a function.}
\label{fig:function}
\end{figure}

\subsubsection{Testing}
The plug-in uses JUnit to run the user's test case on the generated testable function. Both the JUnit test case and the testable function are combined into a compilable class file, and this code is run in a separate process. Running the code in a separate process allows the plug-in to effectively kill the process if it times out, for example if the code contains an infinite loop. If code runs without errors and passes the test case, the snippet is marked as passing, and ranked above all non-passing snippets.

\section{Evaluation}
NLP2TestableCode was evaluated against 47 sample tasks, referenced in the appendix. These tasks are a subset of the total 101 tasks from the original NLP2Code user study \cite{campbell_treude_2017} for which users used NLP2Code's auto-complete feature. We chose to evaluate NLP2TestableCode against the same set of tasks because they are a representation of the types of tasks real users would find helpful.

\subsection{How many code snippets can our approach retrieve?}
In order to maximise the number of retrieved snippets per task, different processing techniques for keywords were compared. The initial number of retrieved snippets without the use of lemmatization and removal of stop words was 2,832. We also measured the effects of both lemmatisation using CoreNLP and stemming using the Porter stemming algorithm \cite{porter}. Stemming is the process of removing word endings, such as "-ing", "-ed" and "s", to reduce a word to a common form, or its stem. Unlike lemmatisation, stemming does not analyse word context or use dictionary look ups, meaning lemmatisation typically outperforms stemming. However, it has been noted that tools like CoreNLP can misinterpret technical language, like that used when discussing software development~\cite{7962368}, because they are trained on more general data. For this reason, comparing both results is necessary. 

\begin{table}[h]
\caption{Comparison of total retrieved snippets.}
  \begin{tabular}{@{}R{1.5cm}rrr@{}}
   \toprule
   \textbf{Omit Stop Words? }& \textbf{No Processing} & \textbf{Stemming} &\textbf{ Lemmatisation} \\
   \midrule
\textbf{No} & 2832 & 4100 & 5091 \\
\textbf{Yes} & 3464 & 5646 & 6954 \\
  \bottomrule
\end{tabular}
  \label{tab:searchresults}
\end{table}

Table \ref{tab:searchresults} shows the effect different keyword processing techniques(no processing, stemming and lemmatisation) and the omission of stop words have on the total number of code snippets retrieved from the Stack Overflow database. The initial 2,832 snippets could be increased to 4,100 using stemming and 5,091 using lemmatisation. The use of stop words further increases the number of retrieved snippets for all processing techniques, with the highest number of snippets being 6,954 using lemmatisation and omitting stop words, with at least one code snippet for all 47 tasks. These results show that despite its limitations, lemmatisation is still more effective than stemming on programming related tasks, and based on these results we chose to implement it within NLP2TestableCode.

\subsection{How many code snippets are compilable?}
We measured the number of compilable snippets before any changes in order to provide a benchmark for the results of code corrections. Each retrieved code snippet was inserted into an empty class and main function before being compiled. Because the results of compiling snippets are dependant on a user's existing code, we chose a simple case like this to represent inserting a snippet into some existing structure while avoiding errors caused by, for example, duplicated pre-existing variables.

\begin{figure}[h]
    \centering
    \includegraphics[width=0.8\linewidth]{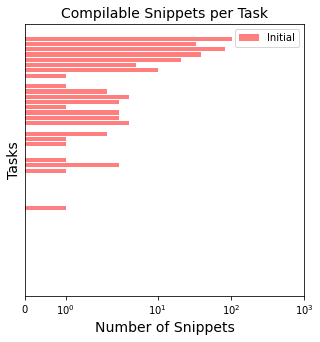}
    \caption{The initial number of compilable snippets per task.}
    \label{fig:initial}
\end{figure}

The number of snippets that compile without changes was 327, out of 6,954 total code snippets. For all 47 tasks, 24 have at least a single compilable snippet. The per task breakdown of compilable snippets can be seen in Figure \ref{fig:initial}, with tasks sorted in descending order based on their final number of compilable snippets after correction.

\subsection{What are the most common error types?}

\begin{figure}[h]
\centering
\includegraphics[width=0.89\linewidth]{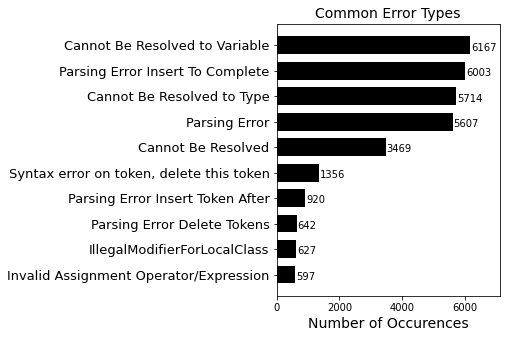}
\caption{10 most common compiler errors in snippets.}
\label{fig:errorTypes}
\end{figure}

We compiled the initial set of code snippets and recorded the types of error codes generated, along with the number of occurrences per error. Each error code generated by the Eclipse compiler corresponds to a constant variable in the Eclipse \texttt{IProblem} interface that provides a short description of the error. This information was used to inform what errors should be the focus of our targeted fixes, in an effort to maximise their effect.

Figure \ref{fig:errorTypes} shows the 10 most common error types generated during the compilation attempt on the initial unmodified set of snippets. Many of these are parsing errors, such as missing semi-colons or incorrectly placed elements, while others include undeclared variables and types. The non-specific `parsing error' and `cannot be resolved' errors make up a large portion of errors and generate compiler messages indicating that these are used when a more specific error cannot be found, for example, the `cannot be resolved' error can be triggered by both variables and type names, likely when this is ambiguous.

\subsection{How many code snippet can our approach make compilable?}
In order to determine the effects of each fix, the number of compilable snippets were recorded after each stage of code correction. Again, code snippets were inserted into an empty main function within an empty class before being compiled.

\begin{figure}[h]
\centering
\includegraphics[width=0.8\linewidth]{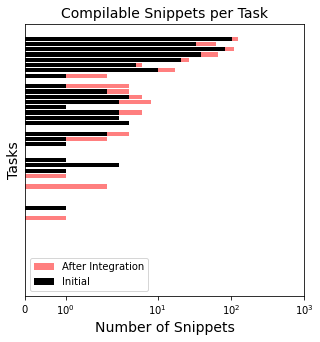}
\caption{The per task breakdown of compilable snippets after integration in pink, compared to the initial number in black.}
\label{fig:integrate}
\end{figure}

\subsubsection{Integration}

Figure \ref{fig:integrate} shows the improvement in number of compilable snippets after the integration step. The total number of compilable snippets was increased from the initial 327 to 470. This is considerable considering how limited the integration tools within NLP2TestableCode are; currently we only handle moving import statements to the top of the file and removing empty classes and duplicate main functions. These results indicate that the use of these structures surrounding example code is common enough that 'snippetising', or reducing unnecessary classes and functions down to their containing statements, can have a non-small effect on the number of compilable snippets.

\subsubsection{Targeted Fixes}

\begin{figure}[h]
\centering
\includegraphics[width=0.8\linewidth]{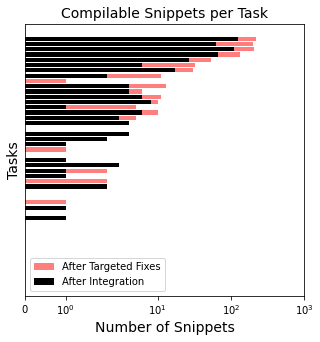}
\caption{Compilable snippets after targeted fixes compared to after integration.}
\label{fig:targeted}
\end{figure}

The use of targeted fixes alongside integration increased the number of compilable snippets from 470 to 968. The per task breakdown can be seen in Figure \ref{fig:targeted}, compared to the number of compilable snippets after integration only. After these fixes, the total number of errors fell considerably, from initially 34,427 errors and 34,002 after the integration step, to 21,514 errors.

\subsubsection{Line Deletion}

Different line deletion configurations were tested, based on the order of line deletion, number of loops over the snippet and acceptance criteria, to determine which configuration maximised the number of compilable snippets. The order of deletion can impact results because often a line has dependencies on other parts of code; for example, deleting a variable declaration before deleting usage or assignment of this variable will generate errors. Similarly, only accepting deletions when they reduce the number of compiler errors, compared to a less-strict acceptance criteria of no increase in errors, can change results and the number of deleted lines considerably.

\begin{figure}[h]
\centering
\includegraphics[width=0.87\linewidth]{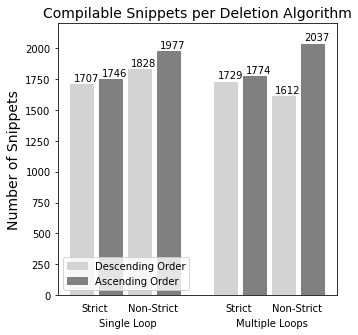}
\caption{Compilable snippets per deletion algorithm}
\label{fig:deletionalgs}
\end{figure}

Figure \ref{fig:deletionalgs} shows the results for each of eight different deletion algorithms based on three options; single or multiple loops over the snippet, strict or non-strict acceptance and descending or ascending order of line deletion. These results exclude empty snippets. In all cases, deleting lines from the bottom up results in more snippets than the same configuration with descending order deletion. In most cases the multiple loop options and non-strict options outperform their alternatives, besides from the descending order, non-strict algorithm which results in 1,612 compilable snippets compared to 1,828 for the single loop alternative and 1,729 snippets for the strict alternative. The non-strict, descending order, multiple loop algorithm outperforms all other algorithms and, because of this, is the algorithm implemented in NLP2TestableCode.

\begin{figure}[h]
\centering
\includegraphics[width=0.8\linewidth]{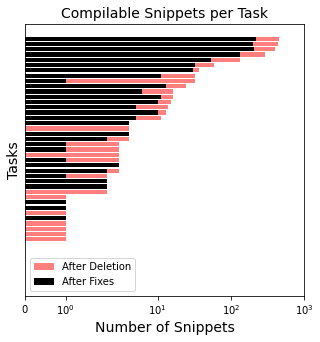}
\caption{The per task increase in compilable snippets after deletion.}
\label{fig:finalsnippets}
\end{figure}

The final number of compilable snippets after deletion is 2,037, a 522.9\% increase from the initial number of compilable snippets. Figure \ref{fig:finalsnippets} shows the final per task break down of compilable snippets compared to only integration and targeted fixes, with 39 out of 47 tasks having at least one compilable snippet.

\subsection{How many code snippets can our approach generate type suggestions for?}

All compilable snippets were run through the type suggestion process. NLP2TestableCode was able to generate at least one type suggestion for 20 tasks, and type suggestions for 316 snippets. This means that at least 20 out of 47 tasks are testable, 51.3\% of the 39 tasks with at least one compilable snippet. User supplied tests may make this number larger as the algorithm that searches for matching variables is less strict; for example, the type suggestion algorithm requires at least one argument. Examples of the types of suggestions generated for sample tasks are listed in Table \ref{tab:types}. In these cases, the algorithm was capable of generating appropriate suggestions.

\begin{table}[h]
\caption{Example of type suggestions generated for tasks.}
  \begin{tabular}{@{}lll@{}}
   \toprule
	\textbf{Task} & \textbf{Arguments} & \textbf{Return} \\
   \midrule
	split string by whitespaces & String & String[] \\
    convert string to integer & String & int \\
    convert uppercase to lowercase & char & char \\
  \bottomrule
\end{tabular}
  \label{tab:types}
\end{table}

\section{Conclusions and Future Work}
NLP2TestableCode's results are promising for the future of automated code reuse. The approaches investigated may be limited, but their effect on the number of code snippets that could be made to integrate is considerable. NLP2TestableCode shows that there is a large amount of improvement in the quality of SO snippets that can be achieved through simple fixes. In our evaluation over 47 tasks, we found that the number of snippets that compile when inserted into an existing piece of code could be increased from 327 to 2,037, and that for 51.3\% of tasks, at least one type suggestion could be generated.

Based on our results, more comprehensive code correction and integration tools would likely only serve to further the number of compilable snippets. There is also room to improve the snippet search algorithm, as currently we only map task keywords to Stack Overflow question titles.

Currently the type suggestion algorithm is limited in that it must find at least one argument type and one return type -- we cannot test void functions or functions without arguments. In addition to this, the type suggestion algorithm often suggests too many argument types, because it will select all variables besides the last one, which is used for the return. It could also be interesting to look at natural language information within tasks to determine testing types. Similarly, we could investigate better ways to automatically generate JUnit test cases and their automatic input and output values.

In the future, evaluating NLP2TestableCode through a user study similar to the one preformed for NLP2Code, would allow us to validate if the new features are useful to developers. This could include evaluating the usefulness of aspects like the highest ranked snippets, the automatic changes made to snippets, type suggestions, automatically generated test cases and the testing process.

\begin{acks}
This research is supported by an Australian Government Research Training Program (RTP) Scholarship.
\end{acks}

\bibliographystyle{ACM-Reference-Format}
\bibliography{ref}


\begin{thebibliography}{25}

%%% the Web address.

\ifx \showCODEN    \undefined \def \showCODEN     #1{\unskip}     \fi
\ifx \showDOI      \undefined \def \showDOI       #1{#1}\fi
\ifx \showISBNx    \undefined \def \showISBNx     #1{\unskip}     \fi
\ifx \showISBNxiii \undefined \def \showISBNxiii  #1{\unskip}     \fi
\ifx \showISSN     \undefined \def \showISSN      #1{\unskip}     \fi
\ifx \showLCCN     \undefined \def \showLCCN      #1{\unskip}     \fi
\ifx \shownote     \undefined \def \shownote      #1{#1}          \fi
\ifx \showarticletitle \undefined \def \showarticletitle #1{#1}   \fi
\ifx \showURL      \undefined \def \showURL       {\relax}        \fi
% invisible to TeX
\providecommand\bibfield[2]{#2}
\providecommand\bibinfo[2]{#2}
\providecommand\natexlab[1]{#1}
\providecommand\showeprint[2][]{arXiv:#2}

\bibitem[\protect\citeauthoryear{Baltes and Diehl}{Baltes and Diehl}{2019}]%
        {Baltes2019}
\bibfield{author}{\bibinfo{person}{Sebastian Baltes} {and}
  \bibinfo{person}{Stephan Diehl}.} \bibinfo{year}{2019}\natexlab{}.
\newblock \showarticletitle{Usage and attribution of Stack Overflow code
  snippets in GitHub projects}.
\newblock \bibinfo{journal}{\emph{Empirical Software Engineering}}
  \bibinfo{volume}{24}, \bibinfo{number}{3} (\bibinfo{date}{01 Jun}
  \bibinfo{year}{2019}), \bibinfo{pages}{1259--1295}.
\newblock
\showISSN{1573-7616}


\bibitem[\protect\citeauthoryear{Barr, Harman, Jia, Marginean, and Petke}{Barr
  et~al\mbox{.}}{2015}]%
        {Barr:2015:AST:2771783.2771796}
\bibfield{author}{\bibinfo{person}{Earl~T. Barr}, \bibinfo{person}{Mark
  Harman}, \bibinfo{person}{Yue Jia}, \bibinfo{person}{Alexandru Marginean},
  {and} \bibinfo{person}{Justyna Petke}.} \bibinfo{year}{2015}\natexlab{}.
\newblock \showarticletitle{Automated Software Transplantation}. In
  \bibinfo{booktitle}{\emph{Proceedings of the 2015 International Symposium on
  Software Testing and Analysis}} \emph{(\bibinfo{series}{ISSTA 2015})}.
  \bibinfo{publisher}{ACM}, \bibinfo{pages}{257--269}.
\newblock
\showISBNx{978-1-4503-3620-8}


\bibitem[\protect\citeauthoryear{Brandt, Dontcheva, Weskamp, and
  Klemmer}{Brandt et~al\mbox{.}}{2010}]%
        {Brandt:2010:EPI:1753326.1753402}
\bibfield{author}{\bibinfo{person}{Joel Brandt}, \bibinfo{person}{Mira
  Dontcheva}, \bibinfo{person}{Marcos Weskamp}, {and} \bibinfo{person}{Scott~R.
  Klemmer}.} \bibinfo{year}{2010}\natexlab{}.
\newblock \showarticletitle{Example-centric Programming: Integrating Web Search
  into the Development Environment}. In \bibinfo{booktitle}{\emph{Proceedings
  of the SIGCHI Conference on Human Factors in Computing Systems}}
  \emph{(\bibinfo{series}{CHI '10})}. \bibinfo{publisher}{ACM},
  \bibinfo{pages}{513--522}.
\newblock
\showISBNx{978-1-60558-929-9}


\bibitem[\protect\citeauthoryear{{Campbell} and {Treude}}{{Campbell} and
  {Treude}}{2017}]%
        {campbell_treude_2017}
\bibfield{author}{\bibinfo{person}{B.~A. {Campbell}} {and} \bibinfo{person}{C.
  {Treude}}.} \bibinfo{year}{2017}\natexlab{}.
\newblock \showarticletitle{NLP2Code: Code Snippet Content Assist via Natural
  Language Tasks}. In \bibinfo{booktitle}{\emph{2017 IEEE International
  Conference on Software Maintenance and Evolution (ICSME)}}.
  \bibinfo{publisher}{IEEE}, \bibinfo{pages}{628--632}.
\newblock


\bibitem[\protect\citeauthoryear{Cottrell, Walker, and Denzinger}{Cottrell
  et~al\mbox{.}}{2008}]%
        {Cottrell:2008:JTS:1370175.1370194}
\bibfield{author}{\bibinfo{person}{Rylan Cottrell}, \bibinfo{person}{Robert~J.
  Walker}, {and} \bibinfo{person}{J\"{o}rg Denzinger}.}
  \bibinfo{year}{2008}\natexlab{}.
\newblock \showarticletitle{Jigsaw: A Tool for the Small-scale Reuse of Source
  Code}. In \bibinfo{booktitle}{\emph{Companion of the 30th International
  Conference on Software Engineering}} \emph{(\bibinfo{series}{ICSE Companion
  '08})}. \bibinfo{publisher}{ACM}, \bibinfo{pages}{933--934}.
\newblock
\showISBNx{978-1-60558-079-1}


\bibitem[\protect\citeauthoryear{Exchange}{Exchange}{2020}]%
        {SEstats}
\bibfield{author}{\bibinfo{person}{Stack Exchange}.}
  \bibinfo{year}{2020}\natexlab{}.
\newblock \bibinfo{title}{All Sites - Stack Exchange}.
\newblock
\newblock
\urldef\tempurl%
\url{https://stackexchange.com/sites?view=list#traffic}
\showURL{%
Retrieved March 10, 2020 from \tempurl}


\bibitem[\protect\citeauthoryear{Foundation}{Foundation}{2020}]%
        {quickfix}
\bibfield{author}{\bibinfo{person}{Eclipse Foundation}.}
  \bibinfo{year}{2020}\natexlab{}.
\newblock \bibinfo{title}{Quick Fix and Quick Assist}.
\newblock
\newblock
\urldef\tempurl%
\url{https://help.eclipse.org/2020-03/topic/org.eclipse.jdt.doc.user/concepts/concept-quickfix-assist.htm?cp=1_2_5}
\showURL{%
Retrieved April 16, 2020 from \tempurl}


\bibitem[\protect\citeauthoryear{JavaParser}{JavaParser}{2019}]%
        {javaParser}
\bibfield{author}{\bibinfo{person}{JavaParser}.}
  \bibinfo{year}{2019}\natexlab{}.
\newblock \bibinfo{title}{JavaParser}.
\newblock
\newblock
\urldef\tempurl%
\url{https://javaparser.org/}
\showURL{%
Retrieved March 27, 2020 from \tempurl}


\bibitem[\protect\citeauthoryear{{Lim}}{{Lim}}{1994}]%
        {wclim}
\bibfield{author}{\bibinfo{person}{W.~C. {Lim}}.}
  \bibinfo{year}{1994}\natexlab{}.
\newblock \showarticletitle{Effects of reuse on quality, productivity, and
  economics}.
\newblock \bibinfo{journal}{\emph{IEEE Software}} \bibinfo{volume}{11},
  \bibinfo{number}{5} (\bibinfo{date}{Sep.} \bibinfo{year}{1994}),
  \bibinfo{pages}{23--30}.
\newblock
\showISSN{0740-7459}


\bibitem[\protect\citeauthoryear{Manning, Surdeanu, Bauer, Finkel, Bethard, and
  McClosky}{Manning et~al\mbox{.}}{2014}]%
        {core}
\bibfield{author}{\bibinfo{person}{Christopher~D. Manning},
  \bibinfo{person}{Mihai Surdeanu}, \bibinfo{person}{John Bauer},
  \bibinfo{person}{Jenny Finkel}, \bibinfo{person}{Steven~J. Bethard}, {and}
  \bibinfo{person}{David McClosky}.} \bibinfo{year}{2014}\natexlab{}.
\newblock \showarticletitle{The {Stanford} {CoreNLP} Natural Language
  Processing Toolkit}. In \bibinfo{booktitle}{\emph{Association for
  Computational Linguistics (ACL) System Demonstrations}}.
  \bibinfo{publisher}{The Association for Computer Linguistics},
  \bibinfo{pages}{55--60}.
\newblock


\bibitem[\protect\citeauthoryear{Martin}{Martin}{2006}]%
        {porter}
\bibfield{author}{\bibinfo{person}{Porter Martin}.}
  \bibinfo{year}{2006}\natexlab{}.
\newblock \bibinfo{title}{Porter Stemming Algorithm}.
\newblock
\newblock
\urldef\tempurl%
\url{https://tartarus.org/martin/PorterStemmer/}
\showURL{%
\tempurl}


\bibitem[\protect\citeauthoryear{Nair, Agrawal, Chen, Fu, Mathew, Menzies,
  Minku, Wagner, and Yu}{Nair et~al\mbox{.}}{2018}]%
        {nair2018dse}
\bibfield{author}{\bibinfo{person}{Vivek Nair}, \bibinfo{person}{Amritanshu
  Agrawal}, \bibinfo{person}{Jianfeng Chen}, \bibinfo{person}{Wei Fu},
  \bibinfo{person}{George Mathew}, \bibinfo{person}{Tim Menzies},
  \bibinfo{person}{Leandro Minku}, \bibinfo{person}{Markus Wagner}, {and}
  \bibinfo{person}{Zhe Yu}.} \bibinfo{year}{2018}\natexlab{}.
\newblock \showarticletitle{Data-Driven Search-Based Software Engineering}. In
  \bibinfo{booktitle}{\emph{Proceedings of the 15th International Conference on
  Mining Software Repositories}} (Gothenburg, Sweden)
  \emph{(\bibinfo{series}{MSR ’18})}. \bibinfo{publisher}{ACM},
  \bibinfo{address}{New York, NY, USA}, \bibinfo{pages}{341–352}.
\newblock
\showISBNx{9781450357166}


\bibitem[\protect\citeauthoryear{Omran and Treude}{Omran and Treude}{2017}]%
        {7962368}
\bibfield{author}{\bibinfo{person}{Fouad Nasser A~Al Omran} {and}
  \bibinfo{person}{Christoph Treude}.} \bibinfo{year}{2017}\natexlab{}.
\newblock \showarticletitle{Choosing an NLP Library for Analyzing Software
  Documentation: A Systematic Literature Review and a Series of Experiments}.
  In \bibinfo{booktitle}{\emph{Proceedings of the 14th International Conference
  on Mining Software Repositories}} \emph{(\bibinfo{series}{MSR ’17})}.
  \bibinfo{publisher}{IEEE}, \bibinfo{pages}{187–197}.
\newblock
\showISBNx{9781538615447}


\bibitem[\protect\citeauthoryear{Overflow}{Overflow}{2020}]%
        {SO}
\bibfield{author}{\bibinfo{person}{Stack Overflow}.}
  \bibinfo{year}{2020}\natexlab{}.
\newblock \bibinfo{title}{Stack Overflow - Where Developers Learn, Share, \&
  Build Careers}.
\newblock
\newblock
\urldef\tempurl%
\url{https://stackoverflow.com}
\showURL{%
Retrieved April 8, 2020 from \tempurl}


\bibitem[\protect\citeauthoryear{Ponzanelli, Bacchelli, and Lanza}{Ponzanelli
  et~al\mbox{.}}{2013}]%
        {Ponzanelli:2013:SSO:2486788.2486988}
\bibfield{author}{\bibinfo{person}{Luca Ponzanelli}, \bibinfo{person}{Alberto
  Bacchelli}, {and} \bibinfo{person}{Michele Lanza}.}
  \bibinfo{year}{2013}\natexlab{}.
\newblock \showarticletitle{Seahawk: Stack Overflow in the IDE}. In
  \bibinfo{booktitle}{\emph{Proceedings of the 2013 International Conference on
  Software Engineering}} \emph{(\bibinfo{series}{ICSE '13})}.
  \bibinfo{publisher}{IEEE}, \bibinfo{pages}{1295--1298}.
\newblock
\showISBNx{978-1-4673-3076-3}


\bibitem[\protect\citeauthoryear{Ponzanelli, Bavota, Di~Penta, Oliveto, and
  Lanza}{Ponzanelli et~al\mbox{.}}{2014a}]%
        {ponzanelli2014mining}
\bibfield{author}{\bibinfo{person}{Luca Ponzanelli}, \bibinfo{person}{Gabriele
  Bavota}, \bibinfo{person}{Massimiliano Di~Penta}, \bibinfo{person}{Rocco
  Oliveto}, {and} \bibinfo{person}{Michele Lanza}.}
  \bibinfo{year}{2014}\natexlab{a}.
\newblock \showarticletitle{Mining StackOverflow to Turn the IDE into a
  Self-Confident Programming Prompter}. In
  \bibinfo{booktitle}{\emph{Proceedings of the 11th Working Conference on
  Mining Software Repositories}} \emph{(\bibinfo{series}{MSR 2014})}.
  \bibinfo{publisher}{ACM}, \bibinfo{pages}{102–111}.
\newblock
\showISBNx{9781450328630}


\bibitem[\protect\citeauthoryear{Ponzanelli, Bavota, Di~Penta, Oliveto, and
  Lanza}{Ponzanelli et~al\mbox{.}}{2014b}]%
        {ponzanelli2014prompter}
\bibfield{author}{\bibinfo{person}{Luca Ponzanelli}, \bibinfo{person}{Gabriele
  Bavota}, \bibinfo{person}{Massimiliano Di~Penta}, \bibinfo{person}{Rocco
  Oliveto}, {and} \bibinfo{person}{Michele Lanza}.}
  \bibinfo{year}{2014}\natexlab{b}.
\newblock \showarticletitle{Prompter: A self-confident recommender system}. In
  \bibinfo{booktitle}{\emph{2014 IEEE International Conference on Software
  Maintenance and Evolution}}. \bibinfo{publisher}{IEEE},
  \bibinfo{pages}{577--580}.
\newblock


\bibitem[\protect\citeauthoryear{Project}{Project}{2019}]%
        {nltk}
\bibfield{author}{\bibinfo{person}{NLTK Project}.}
  \bibinfo{year}{2019}\natexlab{}.
\newblock \bibinfo{title}{Natural Language Tool Kit}.
\newblock
\newblock
\urldef\tempurl%
\url{https://www.nltk.org/}
\showURL{%
Retrieved March 27, 2020 from \tempurl}


\bibitem[\protect\citeauthoryear{Proksch, Bauer, and Murphy}{Proksch
  et~al\mbox{.}}{2015}]%
        {Proksch2015build}
\bibfield{author}{\bibinfo{person}{Sebastian Proksch},
  \bibinfo{person}{Veronika Bauer}, {and} \bibinfo{person}{Gail~C. Murphy}.}
  \bibinfo{year}{2015}\natexlab{}.
\newblock \bibinfo{booktitle}{\emph{How to Build a Recommendation System for
  Software Engineering}}.
\newblock \bibinfo{publisher}{Springer}, \bibinfo{pages}{1--42}.
\newblock


\bibitem[\protect\citeauthoryear{Terragni, Liu, and Cheung}{Terragni
  et~al\mbox{.}}{2016}]%
        {terragni_liu_cheung_2016}
\bibfield{author}{\bibinfo{person}{Valerio Terragni}, \bibinfo{person}{Yepang
  Liu}, {and} \bibinfo{person}{Shing{-}Chi Cheung}.}
  \bibinfo{year}{2016}\natexlab{}.
\newblock \showarticletitle{CSNIPPEX: Automated Synthesis of Compilable Code
  Snippets from Q\&A Sites}. In \bibinfo{booktitle}{\emph{Proceedings of the
  2016 International Symposium on Software Testing and Analysis, {ISSTA}
  2016}}. \bibinfo{publisher}{ACM}, \bibinfo{pages}{118--129}.
\newblock


\bibitem[\protect\citeauthoryear{user}{user}{2020}]%
        {example}
\bibfield{author}{\bibinfo{person}{Unknown user}.}
  \bibinfo{year}{2020}\natexlab{}.
\newblock \bibinfo{title}{How do I convert a String to an int in Java?}
\newblock
\newblock
\urldef\tempurl%
\url{https://stackoverflow.com/questions/5585779/how-do-i-convert-a-string-to-an-int-in-java}
\showURL{%
Retrieved March 27, 2020 from \tempurl}


\bibitem[\protect\citeauthoryear{Yang, Hussain, and Lopes}{Yang
  et~al\mbox{.}}{2016}]%
        {Yang:2016:QUC:2901739.2901767}
\bibfield{author}{\bibinfo{person}{Di Yang}, \bibinfo{person}{Aftab Hussain},
  {and} \bibinfo{person}{Cristina~Videira Lopes}.}
  \bibinfo{year}{2016}\natexlab{}.
\newblock \showarticletitle{From Query to Usable Code: An Analysis of Stack
  Overflow Code Snippets}. In \bibinfo{booktitle}{\emph{Proceedings of the 13th
  International Conference on Mining Software Repositories}}
  \emph{(\bibinfo{series}{MSR '16})}. \bibinfo{publisher}{ACM},
  \bibinfo{pages}{391--402}.
\newblock
\showISBNx{978-1-4503-4186-8}


\bibitem[\protect\citeauthoryear{Yao, Weld, Chen, and Sun}{Yao
  et~al\mbox{.}}{2018}]%
        {StaQC}
\bibfield{author}{\bibinfo{person}{Ziyu Yao}, \bibinfo{person}{Daniel~S. Weld},
  \bibinfo{person}{Wei-Peng Chen}, {and} \bibinfo{person}{Huan Sun}.}
  \bibinfo{year}{2018}\natexlab{}.
\newblock \showarticletitle{StaQC: A Systematically Mined Question-Code Dataset
  from Stack Overflow}. In \bibinfo{booktitle}{\emph{Proceedings of the 2018
  World Wide Web Conference}} (Lyon, France) \emph{(\bibinfo{series}{WWW
  ’18})}. \bibinfo{publisher}{International World Wide Web Conferences
  Steering Committee}, \bibinfo{address}{Republic and Canton of Geneva, CHE},
  \bibinfo{pages}{1693–1703}.
\newblock
\showISBNx{9781450356398}
\urldef\tempurl%
\url{https://doi.org/10.1145/3178876.3186081}
\showDOI{\tempurl}


\bibitem[\protect\citeauthoryear{Yin, Deng, Chen, Vasilescu, and Neubig}{Yin
  et~al\mbox{.}}{2018}]%
        {YinP}
\bibfield{author}{\bibinfo{person}{Pengcheng Yin}, \bibinfo{person}{Bowen
  Deng}, \bibinfo{person}{Edgar Chen}, \bibinfo{person}{Bogdan Vasilescu},
  {and} \bibinfo{person}{Graham Neubig}.} \bibinfo{year}{2018}\natexlab{}.
\newblock \showarticletitle{Learning to Mine Aligned Code and Natural Language
  Pairs from Stack Overflow}. In \bibinfo{booktitle}{\emph{Proceedings of the
  15th International Conference on Mining Software Repositories}} (Gothenburg,
  Sweden) \emph{(\bibinfo{series}{MSR ’18})}. \bibinfo{publisher}{Association
  for Computing Machinery}, \bibinfo{address}{New York, NY, USA},
  \bibinfo{pages}{476–486}.
\newblock
\showISBNx{9781450357166}
\urldef\tempurl%
\url{https://doi.org/10.1145/3196398.3196408}
\showDOI{\tempurl}


\bibitem[\protect\citeauthoryear{Zhang, Jain, Khandelwal, Kaushik, Ge, and
  Hu}{Zhang et~al\mbox{.}}{2016}]%
        {bda}
\bibfield{author}{\bibinfo{person}{Hongyu Zhang}, \bibinfo{person}{Anuj Jain},
  \bibinfo{person}{Gaurav Khandelwal}, \bibinfo{person}{Chandrashekhar
  Kaushik}, \bibinfo{person}{Scott Ge}, {and} \bibinfo{person}{Wenxiang Hu}.}
  \bibinfo{year}{2016}\natexlab{}.
\newblock \showarticletitle{Bing Developer Assistant: Improving Developer
  Productivity by Recommending Sample Code}. In
  \bibinfo{booktitle}{\emph{Proceedings of the 2016 24th ACM SIGSOFT
  International Symposium on Foundations of Software Engineering}}
  \emph{(\bibinfo{series}{FSE 2016})}. \bibinfo{publisher}{ACM},
  \bibinfo{pages}{956–961}.
\newblock
\showISBNx{9781450342186}


\end{thebibliography}

\section{Appendix}

The list of 47 tasks is available at \url{https://github.com/Brittany-Reid/nlp2testablecode/tree/master/data/task,id47.txt}

\end{document}